\documentclass[twocolumn,groupedaddress,showpacs,showfootnotes,preprintnumbers]
{revtex4}
\usepackage{graphicx}
\usepackage{bm}% bold math
\usepackage{amsfonts}

\newcommand{\be}{\begin{equation}}
\newcommand{\ee}{\end{equation}}
\newcommand{\bea}{\begin{eqnarray}}
\newcommand{\beas}{\begin{eqnarray*}}
\newcommand{\eea}{\end{eqnarray}}
\newcommand{\eeas}{\end{eqnarray*}}
\newcommand{\ba}{\begin{array}}
\newcommand{\ea}{\end{array}}
\def\ls{\mathrel{\lower4pt\vbox{\lineskip=0pt\baselineskip=0pt
           \hbox{$<$}\hbox{$\sim$}}}}
\def\gs{\mathrel{\lower4pt\vbox{\lineskip=0pt\baselineskip=0pt
           \hbox{$>$}\hbox{$\sim$}}}}
\newcommand{\e}{\textrm{e}}

\def\be{\begin{equation}}
\def\beq\begin{equation}
\def\ee{\end{equation}}
\def\bea{\begin{eqnarray}}
\def\eea{\end{eqnarray}}

\def\beq{\begin{equation}}
\def\eeq{\end{equation}}
\def\beqa{\begin{eqnarray}}
\def\eeqa{\end{eqnarray}}

\newcommand{\bmat}{\left(\begin{array}}
\newcommand{\emat}{\end{array}\right)}
\newcommand{\mx}{\mbox}
\newcommand{\mt}{\mathtt}

\newcommand{\p}{\partial}
\newcommand{\st}{\stackrel}
\newcommand{\al}{\alpha}
\newcommand{\bb}{\beta}
\newcommand{\ga}{\gamma}

\newcommand{\s}{\sigma}
\newcommand{\om}{\omega}
\newcommand{\Om}{\Omega}

\newcommand{\La}{\Lambda}
\newcommand{\ti}{\widetilde}

\newcommand{\gh}{\widehat{g}}

\newcommand{\mh}{\widehat{m}}
\newcommand{\nh}{\widehat{n}}
\newcommand{\Dh}{\widehat{D}}

\newcommand{\gss}{\st{\circ}{g}}
\newcommand{\ms}{\st{\circ}{m}}
\newcommand{\ns}{\st{\circ}{n}}

\newcommand{\Ds}{\st{\circ}{D}}
\newcommand{\2}{\frac{1}{2}}

\newcommand{\ra}{\rightarrow}

\newcommand{\Ra}{\Rightarrow}

\newcommand{\tir}{\ti{\rho}}
\newcommand{\lcdm}{$\La$CDM}
\newcommand{\ie}{$i.e.$}
\newcommand{\Omm}{\Omega_{\rm m}}

\begin{document}

\title{Current Acceleration from Dilaton and Stringy Cold Dark Matter}

\author{
Tirthabir Biswas $^{1}$
\email[Email:]{tirtho@hep.physics.mcgill.ca},
Robert Brandenberger $^{1}$
\email[Email:]{rhb@hep.physics.mcgill.ca}, 
Anupam Mazumdar $^{2}$ 
\email[Email:]{anupamm@nordita.dk}
and Tuomas Multam\"aki $^{2}$
\email[Email:]{tuomas@nordita.dk}}

\affiliation{$^{1}$ Physics Dept., McGill
University, Montreal, Queb\'ec H3A 2T8, Canada}

\affiliation{$^{2}$ NORDITA,
Copenhagen, Blegdamsvej-17, DK-2100, Denmark }

%\date{}
%\maketitle

\begin{abstract}

We argue that string theory has all the ingredients to provide us with
candidates for the cold dark matter and explain the current
acceleration of our Universe. In any generic string compactification
the dilaton plays an important role as it couples to the Standard
Model and other heavy non-relativistic degrees of freedom such as the
string winding modes and wrapped branes, we collectively call them
stringy cold dark matter. These couplings are non-universal which
results in an interesting dynamics for a rolling dilaton. Initially,
its potential can track radiation and matter while beginning to
dominate the dynamics recently, triggering a phase of
acceleration. This scenario can be realized as long as the dilaton
also couples strongly to some heavy modes.  We furnish examples of
such modes. We provide analytical and numerical results and compare
them with the current supernovae result.  This favors certain stringy
candidates.

\end{abstract} 

\preprint{McGill-05/xx.~NORDITA-2005-42.}


%\keywords{Strings, Branes, Cosmology}

\maketitle

%%%%%%%%%%%%%%%%%%%%%%%%%%%%%%%%%%%%%%%%%%%%%%
\section{Introduction and Summary}

If string theory is the right paradigm for quantum gravity then it
must provide us with all the ingredients to make the currently
observed Universe. According to the most recent measurements
\cite{WMAP}, the universe is to a good approximation spatially flat,
and the current energy budget of the Universe consists of about $30\%
$
non-relativistic matter, most of it being weakly interacting and known
as the cold dark matter (CDM), and about $70\%
$ is in the form of an
unknown source of dark energy which is responsible for the current
acceleration. Unless a dominant fraction of them is hidden in dark
astrophysical objects, Standard Model (SM) baryons can only make up a
small fraction of the total energy density.

Obtaining the SM sector from string theory has been a hot pursuit, see
for a review~\cite{Quevedo}. However in making the present Universe it
is equally challenging to seek a candidate for the cold dark matter
which cannot be otherwise explained within the SM~\footnote{The
coherent oscillations of the Peccei-Quinn axion can explain the origin
of CDM within a minimal extension of the SM (see e.g. Ref.~\cite{Kolb}
for a review) which provides a dynamical mechanism for obtaining a
small $\theta_{QCD}$ and thus solves the {\it strong CP problem}.}.
Finally and most importantly, one also has to obtain the vacuum/dark
energy which seems responsible for the current
acceleration~\cite{Supernovae,WMAP}.

Now it is well known that any string and/or supergravity
compactification to four space-time dimensions usually
\cite{veneziano,taylor} leads to a runaway potential for the dilaton
(see \cite{dil-stab} for efforts to stabilize the dilaton).  Let us
study the rolling dilaton dynamics in the light of dilaton's coupling
to various matter fields. It is important to realize that the dilaton
does not have a universal coupling to all the matter degrees of
freedom \cite{polyakov}, and it is this fact which we are going to
exploit in our paper~\footnote{One may wonder that if the dilaton
starts rolling the gauge couplings change, but as was pointed out in
\cite{veneziano}, when the dilaton runs to the strong coupling regime
the gauge couplings can be determined by algebraic quantities such as
the rank and the Casimir invariants of the gauge group.}.  In order
to have a holistic cosmological evolution in our model, we would
require the following ingredients:

\begin{enumerate}

\item 
The CDM candidate should be treated on an equal footing as the SM
baryons in order to explain why $\Omega_{B}/\Omega_{CDM}\sim {\cal
O}(1)$. Therefore, if we assume that the dilaton couples to both the
sectors, then it should have a similar coupling. In summary, both the
SM and the CDM component should couple weakly (or be uncoupled) to the
dilaton, in order to comply with various empirical constraints from
fifth force experiments, from observational bounds on the variation of
the fine structure constant and of Newton's constant, etc.

\item 
In order to explain the current acceleration, we further require
another CDM sector which is strongly coupled to the dilaton, and which
we coin as SCDM. We will see that, by virtue of its coupling to the
dilaton, the SCDM redshifts much more slowly than CDM and therefore
may dominate the matter energy density at late times even if it was
subdominant at early times.  Moreover, if the coupling is strong
enough, it slows down the rolling of the dilaton, causing the
energy-momentum tensor of the dilaton to lead to an accelerated
expansion of the universe.

\end{enumerate}

We will provide examples of both standard CDM and SCDM in the context
of string theory. At first instance it might look ad hoc to introduce
a new form of dark matter which is strongly coupled to the dilaton,
but as we shall see such states are already present in string theory.

Moreover, our approach to explaining dark energy has an obvious
advantage: we do not have to invoke a new scale such as $10^{-33}$~eV
for the mass of a new scalar field introduced in order to explain the
origin of dark energy as in the case in models with a quintessence
field~\cite{Wetterich,Ratra,Steinhardt} (for alternative explanations,
see~\cite{Jaikumar}).  The natural scale in our approach is solely
governed by the string scale. In this respect our model is minimal and
facilitates making connections to string theory. The reason why
acceleration starts at a much lower scale than the Planck scale, as we
argue, can be traced back to having small initial cosmological energy
densities of SCDM string states in the early (post-inflationary)
universe.

We begin our discussion with the string theoretic motivation for CDM
and SCDM (Section 2). We explain how the string and brane degrees of
freedom we make use of (which are the same ones studied in brane gas
cosmology \cite{BV,ABE}, see also~~\footnote{We are interested in the
late time cosmology of string gas unlike the dynamics of stringy epoch
close to the self dual radius, where the interaction rates are
determined by taking into account of the stringy
corrections~\cite{Frey}.})  couple to the dilaton, and give examples
of candidates for SCDM which couple to the dilaton in a way which can
give rise to acceleration. In Sections 3 and 4, we provide analytical
and numerical descriptions, respectively, of the evolution of the four
distinct components in our cosmology, namely radiation, CDM, SCDM and
the dilaton. We also constrain the couplings by fitting to the
supernovae data. In Section 5, we re-address the cosmic acceleration
problem within the string theoretic framework. We conclude by
summarizing and pointing to future research challenges.

%%%%%%%%%%%%%%%%%%%%%%%%%%%%%%%%%%%%%%%%%%%%%%%%%%%%%%%%%%%%%%%
\section{Motivation from String Theory}

\subsection{Dilaton Gravity}

The main aim of this section will be to find two important couplings,
first the dilaton self coupling which will determine the slope of the
runaway potential, and second the coupling of the dilaton to the
stringy degrees of freedom, allowing us to identify stringy candidates
for CDM and SCDM.

Let us start with $10$-dimensional type II string theory whose low
energy bosonic action is given by
\bea \label{typeII}
\hat{S}_{II} \, &=& \, {1 \over {16 \pi {\hat G}}} \int d^{10}x\sqrt{-g} \\ 
&& \bigl[ e^{-2\phi}\left(\hat{R}+4
\p_{\mh}\phi\p^{\mh}\phi-V(\phi)-{1\over 12}H_3^2\right) \nonumber  \\
&&  - \sum_ne^{a_n\phi}{1\over 2n!}F_n^2 \bigr] \,, \nonumber
\eea 
where hatted quantities denote the full higher ($\Dh=10$) dimensional
objects, $\phi$ is the dilaton, $H_3$ is the field strength of the
antisymmetric tensor field, and the field strengths $F_{n}$'s are $n$
forms (where $n=2,4$ in type IIA theory while n=1,3,5 in type IIB
theory).  To be general, we have also added a potential for the
dilaton. Such a potential could result from quantum corrections
\cite{veneziano,taylor}.

We will assume that all shape and volume moduli of the internal
manifold are stabilized. This could occur by making use of the GKP
scenario \cite{gkp} to stabilize the complex structure (shape) moduli
with the help of various form fields, and then invoking the KKLT ideas
\cite{kklt} to stabilize the volume modulus with the help of
non-perturbative processes. Alternatively, but only in heterotic
string theory, we could use ideas from brane gas cosmology
\cite{BV,ABE} to stabilize both volume \cite{Watson2,Subodh1,Subodh2}
and shape \cite{Edna} moduli by means of string states which are
massless at the self-dual radius (see also \cite{Watson1} for a study
of volume stabilization in the string frame which is applicable both
to type II and heterotic string theories). We assume that these moduli
do not play any substantial role in the late time dynamics (this
assumption is supported by the analyses of \cite{Subodh1,Subodh2}),
and henceforth we will ignore them.

Performing the well-known conformal transformation (relating the
string-frame metric on the left hand side to a function of the dilaton
multiplying the Einstein frame metric on the right hand side)
\be
\hat{g}_{\mh\nh}\ra e^{4\phi/(\Ds+2)}\hat{g}_{\mh\nh}\,,
\label{conf-phi}
\ee
(where $\Ds$ is the number of compactified spatial dimensions), we
recover from (\ref{typeII}) an action in the 10-dimensional Einstein
frame
\bea
\hat{S} \, &=& \, \frac{1}{16\pi \hat{G}}\int d^{10}x\sqrt{-\gh} \\
&& \left\{\hat{R}-\2\p_{\mh}\phi\p^{\mh}\phi-2V(\phi)\right\} \, , \nonumber
\label{eq:sugra}
\eea
where the new potential is related to the original one by rescaling:
\begin{equation}
V(\phi)\ra e^{\phi/2}V(\phi)\,.
\end{equation}
We take the full higher dimensional metric to be:
\begin{equation}
\gh_{\mh\nh}=\left( \begin{array}{cc} g_{mn}(x) & 0\\ 0
&r^2\gss_{\ms\ns}(y)\end{array} \right)\,,
\label{eq:metric}
\ee 
where $r$ is the radius (volume) modulus of the compact space and
where we used the symbol ``$\circ$'' to indicate extra dimensional
quantities.

Then, after integrating over the extra dimensions, one
straight-forwardly obtains:
\be 
S \, = \, \frac{M_p^2}{2}\int
d^4x\sqrt{-g}\left[R-\p_{m}\phi\p^{m}\phi-2V(\phi)\right]\,,
\label{eq:4dgravity}
\end{equation}
where we have defined 
\be
M_p^2=M_s^8 r^6\,,
\ee
as the four space-time dimensional Planck mass, and we have also
rescaled $\phi$
\be
\phi\ra {\sqrt\frac{1}{2}}\phi\,,
\label{phi-rescale}
\ee
in order that it has a canonical kinetic term.

As in \cite{tracking}, we now specialize to the case when $V(\phi)$ is
an exponential potential
\be
V(\phi) \, = \, V_0e^{-2\bb\phi} \, . 
\ee 
Indeed, such exponential potentials are common in lower-dimensional
supergravity models and can also arise in string theory from quantum
corrections (both perturbatively, for example involving string loops
\cite{veneziano,polyakov}, or non-perturbatively, like in gaugino
condensation \cite{taylor}).  Although we keep $\bb$ as a free
parameter in our model, ideally one should be able to derive its value
from the origin of the potential, and we discuss an example in Section
5.

%%%%%%%%%%%%%%%%%%%%%%%%%%%%%%%%%%%%%%%%%%%%%%%%%%%%
\subsection{Wrapped Branes as candidates for CDM and SCDM}

Branes and strings can wrap certain cycles in the internal dimensions.
In general, they are heavy objects with mass scales of the order of
the string scale (an exception are the special modes which are
massless at the self-dual radius - however, such modes exist in
heterotic but not in Type II string theory).  The dilaton in general
couples to these objects, and here we determine these couplings. As in
the usual brane-gas cosmology paradigm \cite{BV,ABE}, we assume that
branes have annihilated in the three non-compact directions, allowing
them to expand, and leaving us only with branes wrapping some or all
of the six internal extra dimensions.  Hence we only need to compute
the couplings for $p$-branes with $p= 2\dots 6$, and for the string
winding and momentum modes.

The brane couplings can be derived \cite{inflation,aaron} starting
from the DBI action, which in the string frame reads
\be S_{DBI}=M_s^{p+1}\int d^{p+1}\s e^{-\phi}\sqrt{-\ga}\,, 
\ee 
where $\ga$ is the metric on the brane world sheet labeled by
$\s$. One has to now make conformal transformations (\ref{conf-phi})
followed by dilaton rescaling (\ref{phi-rescale}) to obtain the brane
coupling $\mu_p$ in the Einstein frame:
\be 
S_{DBI} \, = \, 
M_s^{p+1}\int d^{p+1}\s e^{2\mu_p\phi}\sqrt{-\ga}
\ee
with
\be
2\mu_p \, = \, {p-3\over \sqrt{8}} \, .
\label{brane-exponents}
\ee
The wrapped branes of course appear to us, from the large four
space-time dimensional point of view, as point-like objects. Thus,
their number density redshifts as $a^{-3}$, like non-relativistic
dust. In the brane-gas approximation the four dimensional action for
wrapped $p$-branes are therefore given by (see e.g. \cite{Subodh1})
\bea
S_{\mt{brane}} \, &=& \, 
\int d^4x \sqrt{-g}\rho_0e^{2\mu_p\phi}\left(\frac{a}{a_0}
\right)^{-3} \, \nonumber \\ 
&\equiv& \, \int d^4x \sqrt{-g}\ti{\rho}_p\,,
\label{brane-action}
\eea
where $\ti{\rho}_p$ corresponds to the observed four-dimensional
energy densities of the wrapped branes which appear in the Einstein
equations \footnote{In the above, $a(t)$ is the scale factor of the
four large space-time dimensional cosmology, and $a_0$ is the value of
the scale factor at some reference time at which the energy density is
given by $\rho_0$}.  Note that, because of the coupling, if $\mu_p$ is
positive the energy density $\ti{\rho}_p$ redshifts slower than
$a^{-3}$ if the dilaton increases, and this is why these wrapped
branes can act as candidates for SCDM.

By an analogous prescription we can also obtain the four dimensional
action for string winding modes~\footnote{The reason one cannot
straight forwardly substitute $p=1$ in (\ref{brane-exponents}) to get
the exponent for winding strings lies in the fact that the fundamental
string action does not contain any dilaton coupling, unlike the DBI
action for the solitonic branes which does.}, and momentum modes
starting from the Polyakov action \cite{scott}. For the winding modes
we find
\be
2\mu_1 \, = \, {1\over\sqrt{2}}\,,
\label{string}
\ee
while the momentum modes do not couple to the
dilaton at all \cite{scott}: 
\be
\mu_{\mt{mom}} \, = \, 0 \, .
\ee

One immediately realizes that, since the momentum string modes do not
couple to the dilaton, their energy density redshifts in the same way
as baryons, \ie\ as $a^{-3}$, and therefore they can be candidates for
CDM. On the other hand, the coupled winding states can form the
SCDM. However, since we want SCDM to become important (comparable to
CDM) only at the present cosmological epoch, this means that in the
early universe the energy density of the winding states must have been
much smaller than that of the momentum modes. Here, we present a
qualitative argument based on the masses of these objects as to why we
expect this to be the case. We will address the issue more
quantitatively in Section 5.2.

From Eq.~(\ref{brane-exponents}) we can directly read off the masses
of the $p$-branes as
\be
m_p\sim M_s^{p+1}r^p\,,
\ee
where $r$ is the radion field in the Einstein frame. We assume that
the radion is stabilized. Similarly, from the Polyakov action the
masses for the string winding and momentum modes are given by
\be
m_1\sim M_s^2r\,,
\ee
and
\be
m_{\mt{mom}}\sim {1\over r}\,.
\ee
For the purpose of illustration, here we have assumed that, initially, $\phi$ evolves from close to zero, but similar arguements can be made for more general initial conditions.

If the radion is stabilized at a scale slightly lower than the Planck
scale, say $r^{-1} \sim M_{GUT}$, then because of the $r$ dependence
of the different masses, the momentum modes turn out to be the
lightest, while all the winding string modes and branes are much
heavier. Thus, if there was a period of inflation of the three large
spatial dimensions, and if the temperature after reheating is of the
GUT scale, then the reheating process at the end of inflation can
either perturbatively \cite{AFW,DL} or via parametric resonance
\cite{TB,KLS} produce string momentum modes~\footnote{We assume
instantaneous thermalization, which need not be a correct assumption,
for details see~\cite{Allahverdi}.}. The thermal production of the
winding states (which are very heavy compared to the scale of
inflation) is exponentially suppressed.  There are, however, some
mechanisms by which such heavy modes could be produced
\cite{Wimpzillas,Gubser}, although their number densities would be
much smaller than that of the momentum modes. Thus, in the context of
inflationary cosmology, this may yield one way to explain the
difference in the initial post-inflationary energy densities of the
momentum string modes (CDM) and the winding modes (SCDMs). It is
possible, however, that the winding mode production processes after
inflation are too weak. In this case, one must invoke the primordial
pre-inflationary density of SCDM states to determine the late-time
number density: in Section 5 we argue how a period of inflation of
appropriate length can dilute primordial energy density of SCDMs to
make it just right to address the coincidence problem.

If both baryons and CDM momentum states are produced during reheating
after inflation, then it should be expected that their densities are
comparable, as long as the masses of both types of states are small
compared to the typical energy scale during reheating. If the initial
post-inflationary baryon density is not changed substantially by later
stages of baryon production, the number densities of baryons and CDM
particles are expected to be comparable.

Let us then assume that after inflation we have very small energy
densities of the different wrapped brane/string gases. Now observe
that because of the difference in the coupling exponents $\mu_p$'s,
all the species redshift differently. As $\phi$ is increasing, it is
clear that very soon the largest energy density among these different
species would correspond to the one with the largest positive
$\mu_p$. From (\ref{brane-exponents}) and (\ref{string}), one finds
that this corresponds to the 6-branes, with
\be
\mu_6 \, = \, {3\over2\sqrt{8}}\,.
\label{6-exponent}
\ee
Thus, according to our scenario, at late times the dark sector of our
universe will consist of three main components: (a) the dilaton
behaving as dark energy, (b) string momentum modes playing the role of
ordinary CDM and (c) 6-branes, which wrap the six extra dimensions,
playing the role of SCDM. As we shall see in the next section, by
virtue of their coupling (\ref{6-exponent}) to the dilaton, 6-branes
can trigger a late phase of acceleration.

%%%%%%%%%%%%%%%%%%%%%%%%%%%%%%%%%%%%%%%%%%%%%%%%%%%%%%%%%%%%%

\section{Cosmological Model}

\subsection{Acceleration due to Coupling: Basic Mechanism}

We begin by analyzing a simple scalar-tensor theory action for dilaton
gravity, where the dilaton has a run-away potential:
\bea \label{1-action}
S_{\mt{grav}+\phi} \, &=& \, M_p^2\int d^4x\ \sqrt{-g} \\
&& \left[{R\over2}-\frac{(\p\phi)^{2}}{2} - V_0e^{-2\bb\phi}\right]\,.
\nonumber
\eea
In principle, $\phi$ can be identified with any suitable moduli field
as exponential potentials are quite common for other kinds of moduli
fields as well, for example, in flux compactifications
\cite{duff}. However, for the purpose of this paper we  focus on the
dilaton.  As in conventional cosmology, gravity is sourced by the
matter-radiation stress energy tensor which can be derived from an
action of the form
\be 
S_i=\int d^4x\ \sqrt{-g}\tir_i\,, 
\ee 
where the index $i$ represents the different components of
matter/radiation and $ \tir_i$ is the observed energy density that
gravitates. The Hubble equation reads:
\be 
H^2 = {1\over
3}\Big(\sum_i\tir_i+\frac 12 \dot\phi^2+ V(\phi)\Big)\equiv{1\over
3}\Big(\sum_i\tir_i+\tir_\phi\Big)\,.
\label{1-hubble}
\ee
Note there is a departure from the conventional cosmology. We allow
different matter components to couple to $\phi$:
\be 
\tir_i=e^{2\mu_i \phi}\rho_i\,, 
\ee 
where we further assume that the $\phi$-independent ``bare density''
$\rho_i$ behaves like a perfect fluid \ie, it obeys the continuity
equation:
\be 
\dot{\rho_i}+3H(p_i+\rho_i)=0\,,
\label{continuity}
\ee
with an equation of state, $p_i=\om_i \rho_i$.

If a scalar field $\phi$ is to act as a quintessence type field, then
it must be virtually massless, i.e. $m_\phi\sim 10^{-33}ev$.  If
$\phi$ is coupled to SM particles, it mediates a fifth force.  Fifth
force experiments/observations constrain such couplings of $\phi$ to
SM degrees of freedom to be very small,
$\mu_{\mt{SM}}<10^{-4}-10^{-5}$ (see e.g. \cite{Carroll}).  For a
dilatonic $\phi$, a ``least coupling'' principle was proposed by
Damour and Polyakov~\cite{polyakov} which can explain how its
couplings to SM particles can become increasingly small as $\phi$
rolls towards the strong coupling regime.  We note in passing that
observations and experiments on variations of physical constants like
the fine structure and Newton's constants also give us similar bounds
on $\mu_{\mt{SM}}$.  Therefore in the following we will assume that
radiation (r), baryons (b) and CDM (string momentum modes) have
tiny/no couplings to $\phi$.

The dynamics of $\phi$ depends on the exponent $\bb$, and the coupling
of $\phi$ to various matter-radiation components. The equation of
motion for $\phi$ is given by
\be \label{1-Qevolution}
\ddot\phi + 3H\dot\phi \, = \, -V'_{\mt{eff}}(\phi)\,,
\ee
with
\be 
V_{\mt{eff}}(\phi) \, = \, V_0e^{-2\bb\phi}+\sum_{i=m,sc}e^{2\mu_i \phi}\rho_i\,,
\ee
where the sum runs over all matter states. For simplicity, we have
assumed that the gauge bosons do not contribute to an effective
potential for $\phi$. This is strictly true if on average the electric
field and magnetic field vanishes.  There could be other forms of
radiative matter which could induce an effective coupling, as
discussed for example in~\cite{ovrut}.  Weakly coupled matter (stringy candidates for CDM and baryons) are denoted by the subscript $''m''$, and the SCDM by the subscript $''sc''$.

It was argued in~\cite{tracking} that, when the Universe evolves
adiabatically, the dilaton $\phi$ tracks the minimum of the effective
potential which evolves in time with the redshifting of the
$\rho_i$'s. How fast or slow the dilaton is evolving mostly depends on
the equation of state parameter $\om_i$, and on the coupling exponent
$\mu_i$ of the dominating matter component. Thus, at earlier times
when radiation or weakly coupled matter is dominant, $\phi$ rolls fast
and tracks the relevant matter/radiation component, remaining
subdominant. However, once SCDM becomes important, the dilaton slows
down, starts to dominate the energy density, and our universe enters a
phase of acceleration.  Let us now see this in more detail.

%%%%%%%%%%%%%%%%%%%%%%%%%%%%%%%%%%%%%%%%%%%%%%%%%

\subsection{Exact Two-fluid Analysis}

We will first work within the approximation that there are only two
dominating components in the Universe, the dark energy determined by
$\phi$ and a single matter/radiative component. To keep things
general, we do not specify which form of matter it is.  In this
approximation, the Einstein constraint equation (\ref{1-hubble})
determining the Hubble expansion rate reads
\be 
H^2 = {1\over
3}\Big(\e^{2\mu \phi}\rho+\frac 12 \dot\phi^2+ V(\phi)\Big)\,.
\label{twofluid}
\ee
From Eq.~(\ref{1-Qevolution}), it follows that the evolution equation
for $\phi$ is given by
\be
\ddot\phi+3H\dot\phi  =  2(\bb V_0\e^{-2\beta \phi}-\mu\e^{2\mu \phi}\rho)\,.
\ee
Solving the continuity equation (\ref{continuity}), we obtain,
\be
\rho=\rho_{0}\left(\frac{a}{a_0}\right)^{-3(1+\om)}
\ee
where $\rho_0$ is the present day value of the energy density.  

Performing a change of variables $\eta=\ln(a(t))$ leaves us with the
following set of equations:
\bea 
H^2 & = & {1\over 3}{\e^{2\mu \phi}\rho+V_0\e^{-2\bb \phi}}\over 
{1-{1\over 6}\phi^{\prime 2}}
\nonumber\\ 
\phi^{\prime\prime}+\Big(3+{H'\over H}\Big)\phi' & = & 2\beta\frac{V_0}{H^2} 
\e^{-2\bb \phi}-2\mu{\rho\over H^2}\e^{2\mu \phi},\nonumber 
\eea 
where $'\equiv d/d\eta$. The system of equations can be solved
analytically to obtain the ``tracking solution'', the solution where
$\phi(t)$ tracks the minimum of its ``effective'' potential. The
solution is given by
\bea \label{twofluidsol}
&& e^\phi \, = \\
&& \left[{V_0\over\rho}\Big({\beta(\beta+\mu)-3/4(1+\om)\over
\mu(\mu+\beta)+3/4(1+\om)^2-3/8(1+\om)}\Big)\right]^p \nonumber
\eea
where
\be
p \, = \, {1 \over {2(\beta+\mu)}} \, ,
\ee
and
\be
a(t)=a_0\left({t\over t_0}\right)^{2\over3(1+\om_\phi)}\mx{ with }
\om_\phi={-{\mu/\bb-\om\over\mu/\bb+1}} \, .
\label{scale}
\ee
A few remarks are now in order. First, we observe that the energy
densities $\ti{\rho}$ and $\ti{\rho}_\phi$ track each other and
redshift with precisely the same equation of state parameter
$\om_\phi$ that was derived in the adiabatic approximation in
\cite{tracking}:
\bea
\ti{\rho}_\phi \, &=& \, {\mu(\mu+\beta)+3/4(1+\om)^2-3/8(1+\om)\over
\beta(\beta+\mu)-3/4(1+\om)}\ti{\rho} \nonumber \\
&\equiv& \, r\ti{\rho}\sim a^{-3(1+\om_{\phi})}\,.
\label{1-solution}
\eea
One recovers the value of the tracking ratio
\be
 r={\mu\over \bb}\,,
 \label{adiabatic}
\ee
obtained in the adiabatic approximation \cite{tracking} when
$\mu,\bb\geq{\cal O}(1)$ and the terms involving $(1+\om)$ can be
ignored.

Second, we realize that real solutions for $\phi$ only exist if the
numerator on right hand side of (\ref{twofluidsol}) is positive (the
denominator is positive as long as $\om>-1/2$). In other words, in
order for a tracking solution to exist we require:
\be
\beta(\beta+\mu) \, > \, {3(1+\om)\over 4}\,.
\label{cond}
\ee
Since we want such tracking behaviour to hold during the
radiation-dominated phase, substituting $\mu_r=0$ and $\om_r=1/3$ in
Eq.~(\ref{cond}), we find:
\be
\bb \, > \, 1 \,.
\label{beta-bound}
\ee 
We note in passing that, if the condition Eq.~(\ref{cond}) is not
satisfied, then we end up in a different non-tracking late time
attractor solution.  In this case the field does not follow the
minimum but lags more and more behind as the universe expands, and the
energy density becomes dominated by the dilaton from the very
beginning.  From the point of view of cosmology this is an
uninteresting scenario.

Finally, we find from Eqs.~(\ref{scale}) that acceleration happens
when
\be
{\mu_{sc}\over\bb} \, > \, {3\over 2}({1 \over 3} + \om_{sc})\,.
\label{1-accel}
\ee  
For winding brane SCDM ($\om_{sc}=0$), combining (\ref{1-accel}) with
(\ref{beta-bound}), we find that we can account for a late
acceleration phase provided our SCDM has a coupling exponent
\be
\mu_{sc} \, > \, \2 \, .
\ee
Note, in particular, that the exponent which we computed for wrapped
6-branes satisfies this bound.

\subsection{Three fluids system} 

Our exact analysis of two fluids model suggests that, if the dilaton
couples to SCDM, then there can be an interesting dynamics which
triggers a phase of acceleration. In order to study more precisely how
such a transition happens, one has to consider a more realistic
cosmological model where, in addition to the SCDM fluid, we also have
a normal CDM component that is uncoupled or weakly coupled to the
dilaton. Such a model has the advantage that, at early times, the
evolution is standard, i.e.  $H\sim a^{-3/2}$, and, at late times, as
SCDM plus dark energy begin to dominate, we can have accelerated
expansion.

The new system of equations (\ref{1-hubble}-\ref{1-Qevolution})
becomes:
\begin{eqnarray}
\label{Hubble}
H^2 &=&\frac{1}{3M_{p}^2}\left(\ti{\rho}_\phi+\ti{\rho}_{sc}+
\ti{\rho}_m\right)\,,
\\ 
\ddot\phi+3H\dot\phi&=& 2(\bb V_0e^{-2\phi}-\mu_{sc}\rho_{sc}
e^{2\mu_{sc}\phi}-\mu_m \rho_{m}e^{2\mu_m \phi}) \nonumber \\
&=& 2[\bb V(\phi)-\mu_{sc}\ti{\rho}_{sc}-
\mu_m\ti{\rho}_m]\,,
\label{eqm1} 
\end{eqnarray}
  For simplicity, we assume for the time being that $\mu_m=0$; we will see
later that as long as $\mu_m<<\bb$, it hardly makes any difference to
our analysis.  Although no analytical solution to
Eqs.~(\ref{Hubble}-\ref{eqm1}) is found, one can obtain some insight
into the cosmological evolution resulting from these equations by
considering various limiting situations.

{\bf First transition}: First, consider a situation when the energy
density in SCDM is much smaller than that in both ordinary matter and
in $V(\phi)$, so that one can neglect the second terms in
Eqs.~(\ref{Hubble}) and (\ref{eqm1}).  In this case, the equations
from the two fluid discussion apply, making use of $\mu = 0$ and $w =
0$. This is the phase when the dilaton essentially rolls down freely
along its potential $V(\phi)$, tracking the energy density of ordinary
matter \cite{tracking} as given by (\ref{1-solution}):
\be 
\ti{\rho}_\phi \, = \, \frac{3}{8\bb^2-6} \ti{\rho}_m \,.
\label{matter}
\ee
The scale factor $a(t)\propto t^{2/3}$ as in the usual matter
dominated era. Note that during this period the value of $\phi$ which
is the minimum of the ``effective potential'' for $\phi$ (the sum of
$V(\phi)$ and $\ti{\rho}_{sc}$) is redshifting more slowly than the
value of $\phi(t)$ in this phase (this can be seen by going back to
(\ref{twofluidsol}), since it is setting $\mu = \mu_{sc}$ in this
equation which gives the time dependence of the minimum), and thus
$\phi(t)$ eventually catches up with the minimum. After this point,
the value of $\phi(t)$ tracks the minimum. In this case, we know that
the energy density of $\phi$ will track SCDM, as given in
Eq.~(\ref{1-solution}).  It is clear then, that when the right hand
sides of Eqs.~(\ref{1-solution}) and (\ref{matter}) become equal, \ie
when
\be 
\frac{3}{4\bb^2-3}\ti{\rho}_m \, = \, \frac{\mu}{\bb}
\ti{\rho}_{sc}
\ee
and thus
\be
\ti{\rho}_{sc} \, = \, \frac{3\bb}{\mu(4\bb^2-3)} \ti{\rho}_m 
\ee 
a first transition, namely the transition between the period when
$\phi(t)$ is moving freely and when it starts tracking the minimum of
the effective potential (set by the interaction of the dilaton
potential and the contribution from the SCDM term) occurs \footnote{In
the above, we have made use of the adiabatic approximation
(\ref{adiabatic}).}. From this moment on, both $\ti{\rho}_{sc}$ and
$\ti{\rho}_Q$ start to redshift as Eq.~(\ref{1-solution}), with
$\om\ra 0$ and $\mu\ra \mu_{sc}$.
 
{\bf Second Transition}: The second transition \footnote{The exact
sequence of these two events are actually not important for the basic
mechanism to work. However for the relevant values of $\mu/\bb$ the
two events occur in the order described.} corresponds to when the
universe enters a phase of acceleration. This happens approximately
when $\ti{\rho}_d \, \equiv \, \ti{\rho}_{sc}+ \ti{\rho}_\phi \, = \,
(1+r_{sc}) \ti{\rho}_{sc}$ becomes equal to the energy density of the
ordinary matter components, as can be readily seen from the Hubble
rate Eq.~(\ref{Hubble}). Thus, we enter a phase of acceleration when:
\be
\ti{\rho}_{sc} \, = \, {1\over1+r_{sc}}\ti{\rho}_m
\, \approx \, {1\over1 + \mu_{sc}/\bb}\ti{\rho}_m \, ,
\ee 
where in the last step we again made use of the adiabatic
approximation.

First, notice that depending on how large the ratio $\mu_{sc}/\bb$ is,
one needs a very small fraction of SCDM compared to ordinary weakly
coupled matter to initiate the phase of acceleration.  Since SCDM
behaves like non-relativistic dust ($\om_{sc}=0$) it would cluster
around galaxies and contribute to the current dark matter abundance,
thereby changing the baryon to dark matter ratio in recent
times. Since there is reasonable agreement between this ratio as
obtained from the cosmic microwave background (CMB) (which measures
the primordial fluctuations at redshifts of around $z=1100$) and
measurements of rich clusters of galaxies (at recent redshifts), it is
important that this ratio does not change too much in our model. This
is guaranteed if $\ti{\rho}_{sc}$ is only a small fraction of
$\ti{\rho}_m$. As pointed out in \cite{tracking} this scenario of
changing baryon to dark matter abundance offers the intriguing
possibility of reconciling a 10 to 20 percent discrepancy in the
estimates of baryonic abundances coming from CMB and big bang
nucleosynthesis (BBN) measurements \cite{subir}.

{\bf What if SCDM is a radiative fluid?}: So far, we have assumed that
SCDM is a form of dark matter (winding states) but is strongly coupled
to the dilaton. What if it were to behave as a radiative fluid?  For
example, in \cite{ovrut} it was proposed how thermal fluctuations of a
massless field with quartic interactions can act as a radiative fluid
with precisely the kind of exponential coupling to the dilaton that we
considered here.  In this case $\om_{sc}=1/3$, and we will observe it
as a component of dark energy. This is because after the first
transition both SCDM and $\phi$ obey the same equation of state
$\om_\phi$ as computed in Eq.~(\ref{1-solution}), and cosmologically
we cannot distinguish between the two. The total dark energy would be
then given by $ \ti{\rho}_d$. Obviously, in this setup the ratio
between dark matter and baryons remains constant and we do not
encounter any constraints on the ratio $\mu_{sc}/\bb$ on this basis.
From (\ref{1-accel}) we find that in order to have acceleration the
ratio just needs to satisfy \be {\mu\over\bb} \, > \, 1 \, .  \ee

{\bf Acceleration epoch}: In either case, the main constraint comes
from trying to match the observed equation of state, $\om_\phi<-0.76$
\cite{WMAP,Supernovae}. From Eq.~(\ref{scale}) we find such a bound implies
\be 
{\mu_{sc}\over\bb} \, > \, 3 
\ee
for SCDM and
\be \label{1-ratio}
{\mu_{sc}\over\bb} \, > \, 5  
\ee
for radiation. How to get such large value of the ratio from string
theory is discussed in the following section.

For these ratios of $\mu_{sc}/\bb$ one can also estimate approximately
the onset of the period of acceleration. It is the time when
$\ti{\rho}_d\approx\ti{\rho}_m$. Now
\bea
1 \, &=& \, {\ti{\rho}_{d,acc}\over\ti{\rho}_{m,acc}}
\, = \, {\ti{\rho}_{d0}\over\ti
{\rho}_{m0}}\left(a_{acc}\over a_0\right)^{-3\om_Q} \nonumber \\
&=& \, {\Om_{d0}
\over\Om_{m0}}\left({1\over 1+z_{acc}}\right)^{-3\om_Q} \, . 
\eea
Using the known dark matter to dark energy ratio, we find
\be
z_{acc}\sim 0.5-1.0\,, 
\ee 
which is in good agreement with the supernova data.

Finally, we note that most of our analysis of the transitions go
through identically even if $\mu_m$ is non-zero but small. This is
because, if $\mu_m$ is tiny it changes very little the tracking ratio
in the matter phase (\ref{matter}) which enters in the calculation for
the epoch of first transition. Also,
$$\tir_m\sim a^{-3+{\cal O}(\mu_m/\bb)}.$$ Thus, weakly coupled matter
redshifts essentially the same way as in the standard cosmology and
therefore the second transition is also unaffected.

To summarize, we have three distinct cosmological eras depending on
which terms are most dominant on the right hand side of the Hubble
equation Eq.~(\ref{Hubble}):

\begin{enumerate}

\item
Radiation era, when $\ti{\rho}_{sc}$ and $\ti{\rho}_m$ are
negligible while $\ti{\rho}_\phi$ tracks $\ti{\rho}_r$,

\item
Matter era, when $\ti{\rho}_{sc}$ and $\ti{\rho}_r$ are negligible
while $\ti{\rho}_\phi$ tracks $\ti{\rho}_m$, and finally,

\item
an Acceleration phase, where stringy components, viz. $\ti{\rho}_d$
starts to dominate over $\ti{\rho}_m$. 

\end{enumerate}

As far as the parameters of our model are concerned, first we find
that in order to have a late time acceleration phase we need the ratio
$\mu_{sc}/\bb$ to satisfy conditions like Eq.~(\ref{accel}). On the
other hand, to have an earlier non-accelerating phase $\mu_r,\mu_m$
has to be small which is consistent with constraints coming from Fifth
force experiments, and observations of variation of $G_N$ and $\al_f$
which lead to the condition $\mu_{SM}<10^{-5}$.  Finally, the
requirement that $\phi$ should track radiation in the early universe
tells us that $\bb>1$.  In spite of these constraints, there is a huge
parameter space where we seem to agree with the cosmological and
particle physics observations, which is encouraging. Let us therefore
study this model in more detail using numerical techniques and see
whether the allowed parameter space can be made more precise.

%%%%%%%%%%%%%%%%%%%%%%%%%%%%%%
%%%%%%%%%%%%%%%%%%%%%%%%%%%%%%%%%%%%%%%%%%%%%%%%%%%%%%%%%%%%%%%%%
\begin{figure}
\centering
\includegraphics[width=5cm,angle=-90]{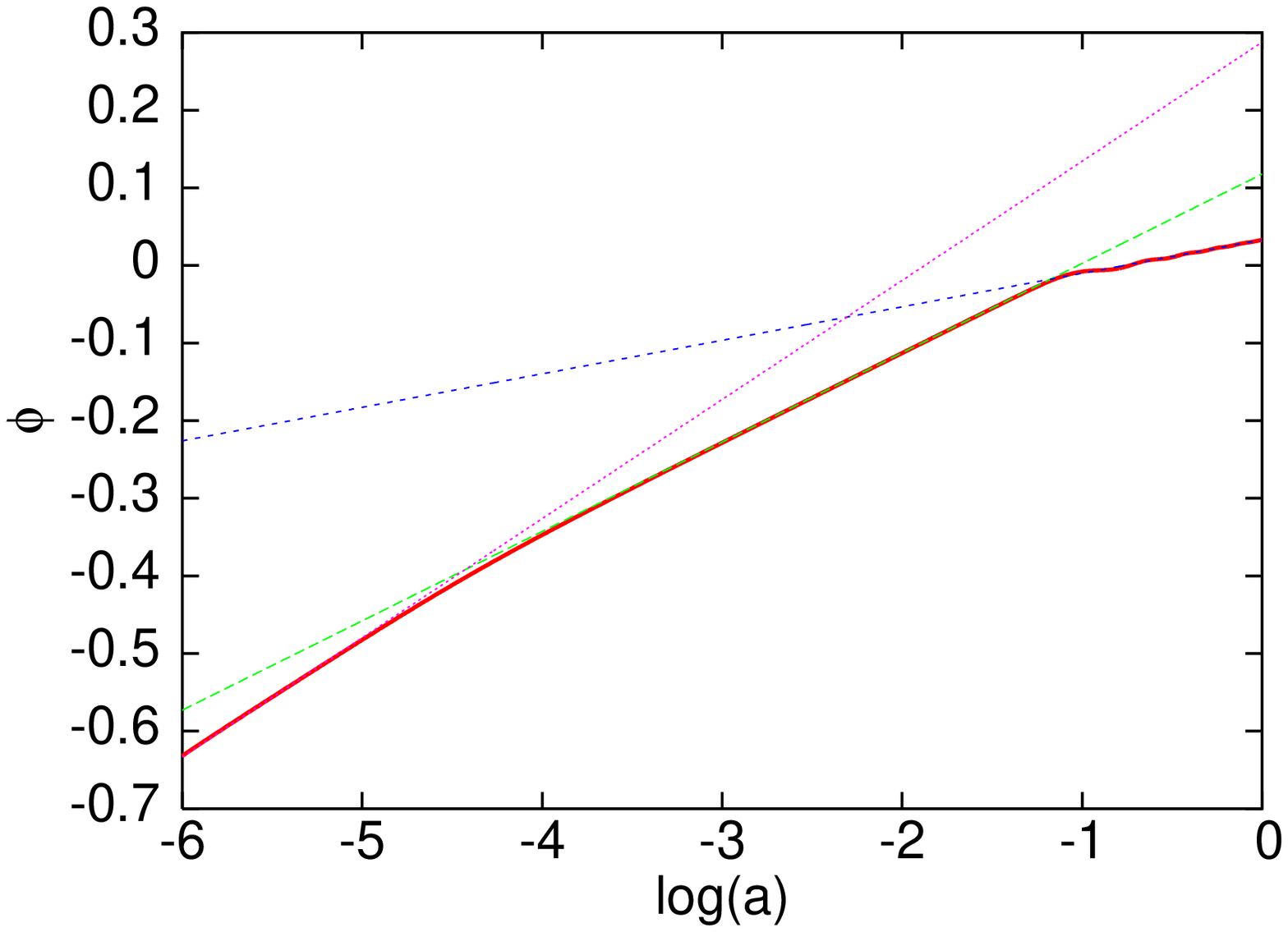}
\caption[fig1]{\label{fig1} The evolution of $\phi$ as a function
of the logarithm of the cosmological scale factor for the full
numerical solution (solid red) and for the approximations to the
solutions corresponding to the evolution at late (blue dashed) times,
early times (green long dashed) time and in the initial
radiation-dominated (purple dot-dashed) phase.}
\includegraphics[width=5cm,angle=-90]{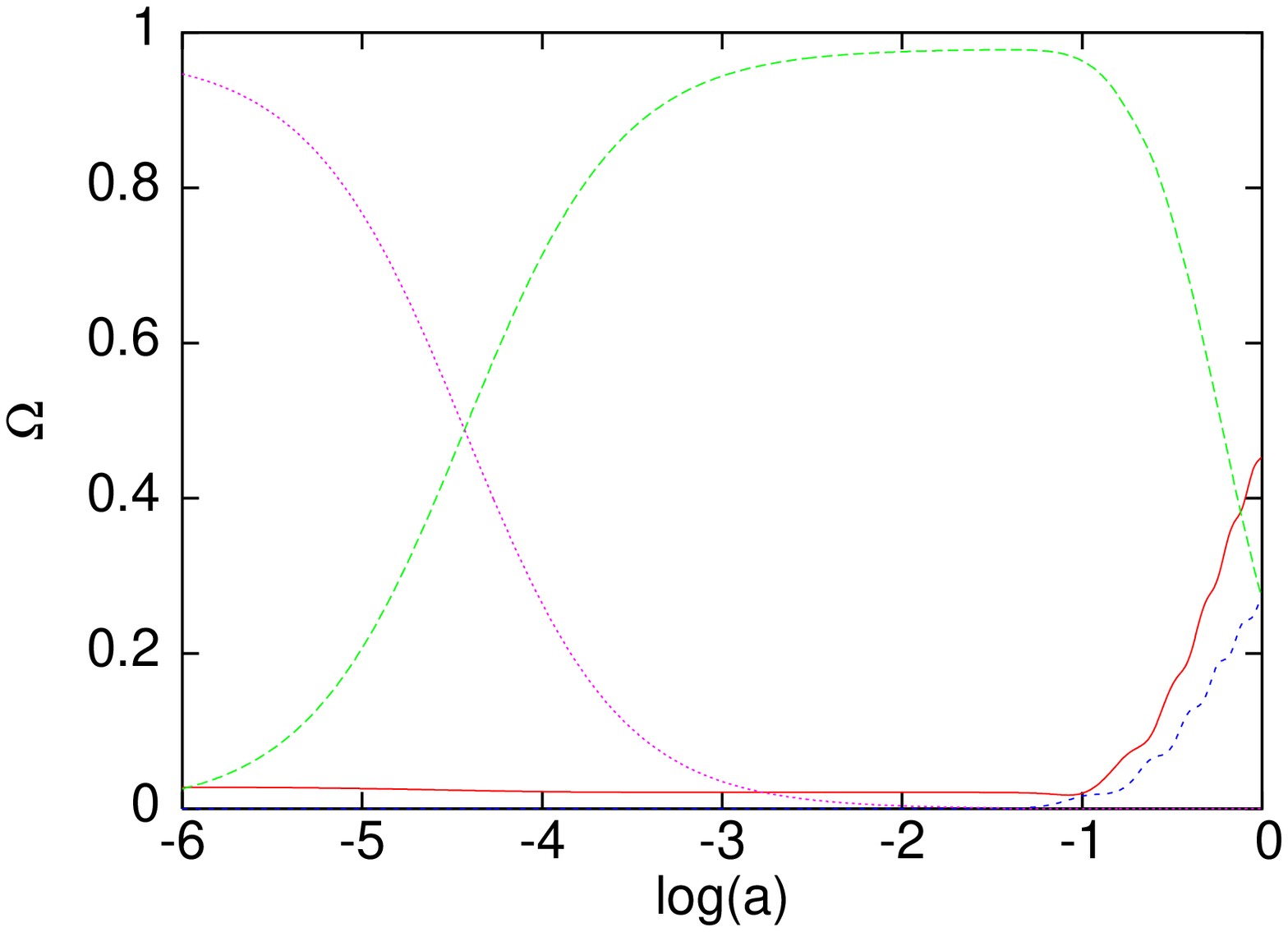}
\caption[fig2]{\label{fig2} The relative energy densities in the
$\phi$ field (red solid), in coupled CDM (SCDM)(blue dashed),
uncoupled CDM (green long dashed) and in radiation (purple dot
dashed). }
\end{figure}

%%%%%%%%%%%%%%%%%%%%%%%%%%%%%%%%%%%%%%%%%%%%%%%%%%%%%%%%%%%%%%%%%

\section{Numerical work}

\subsection{Evolution and Abundances}

In order to verify our analytical approximations and see the precise
evolution of the system, we solve the three fluid system
numerically. We start deep in the matter dominated era and choose
initial conditions such that the field initially follows the
approximate early time solution. As an example, the evolution of the
field for the set of parameter values $\beta=6,\ \mu=10,\
\Om_{\tilde{m}}=0.27,\ \Omm=0.01$, and $V_0$ fixed by requiring
flatness is shown in Fig.~\ref{fig1}. The plot depicts the field value
as a function of time (measured in terms of the logarithm of the scale
factor). The solid red curve represents the full numerically
determined evolution.  In the same figure we also show the analytical
approximations given by Eqs. (\ref{Hubble}-\ref{eqm1}) in a universe
that is radiation dominated, i.e. $\exp(\phi)=(8\beta^2V_0/3
\rho_r)^{1/2\beta}$ (dot-dashed purple curve), dominated by ordinary
matter, i.e.  ($\exp(\phi)=(8\beta^2V_0/3 \tilde{\rho}_m)^{1/2\beta}$
long dashed green curve), or described by the late time solution
(dashed blue curve) when the field follows the effective minimum of
the potential Eq.~(\ref{1-solution}).  From the figure one can see how
initially the field follows the radiation dominated solution until it
crosses the matter dominated solution, after which the evolution of
the fields is well described by it.  At late times, the field follows
the solution of the evolution of the minimum of the effective
potential. Note also how the crossover from one solution to another is
very rapid and smooth.

The relative contributions $\Omega_i$ of various matter constituents
(labeled by $i$) to the critical density of a spatially flat universe
are shown in Fig.~\ref{fig2}. It is clear how at late times the
coupled dark matter component starts dominating along with the $\phi$
field.

%%%%%%%%%%%%%%%%%%%%%%%%%%%%%%%%%%%%%%%%%%%%%%%%%%%%%%%%%%%%%%%%

\begin{figure}
\centering
\includegraphics[width=5cm,angle=-90]{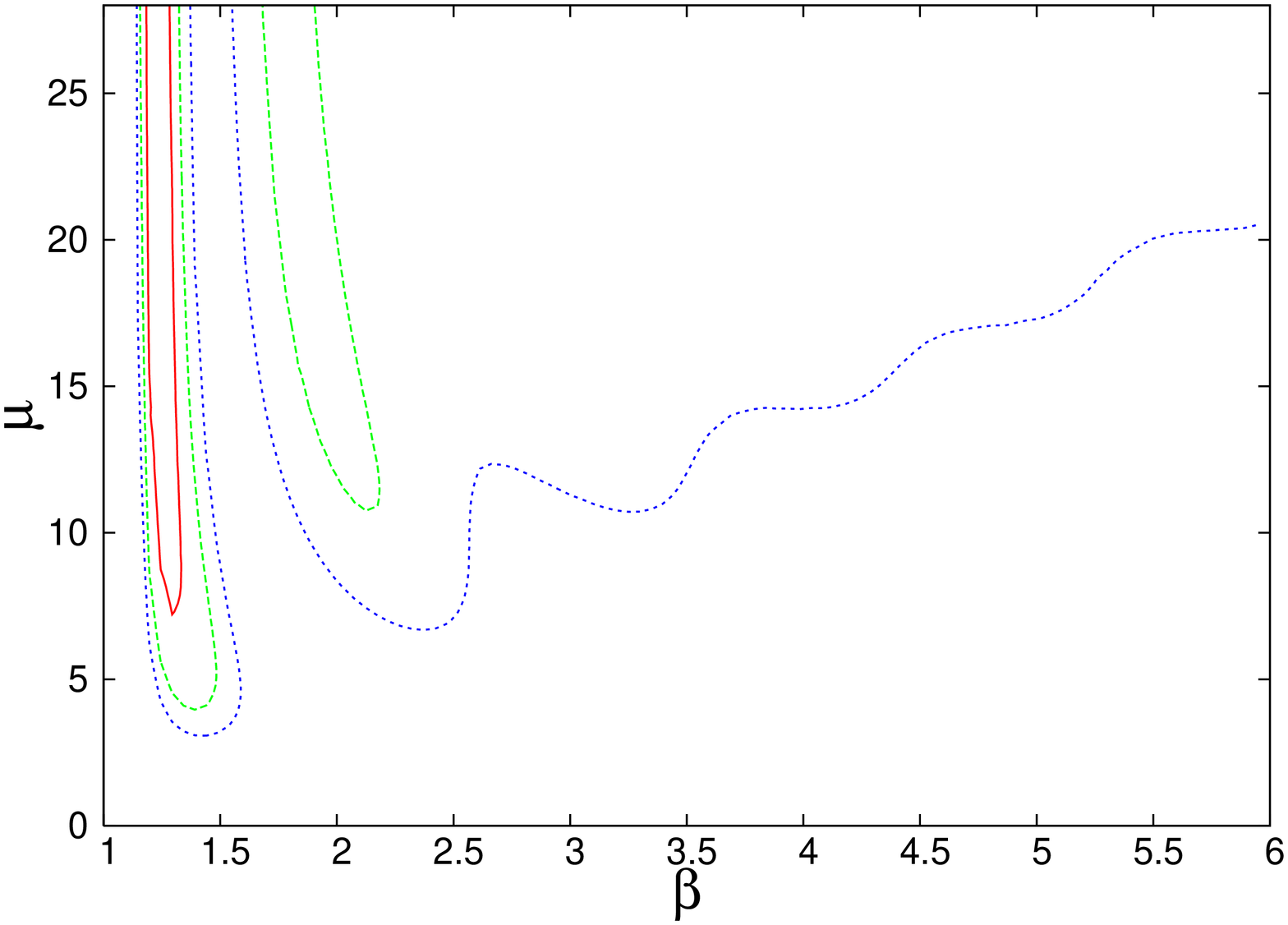}
\caption[fig3]{\label{fig3}Contour plot of $\Delta\chi^2$
($\Omega_{\tilde m}=0.27,~\Omega_m=0.01$). Red (solid) contour
corresponds $-4$, green (long-dashed) to $-2$ and blue (short-dashed)to $0$.}
\includegraphics[width=5cm,angle=-90]{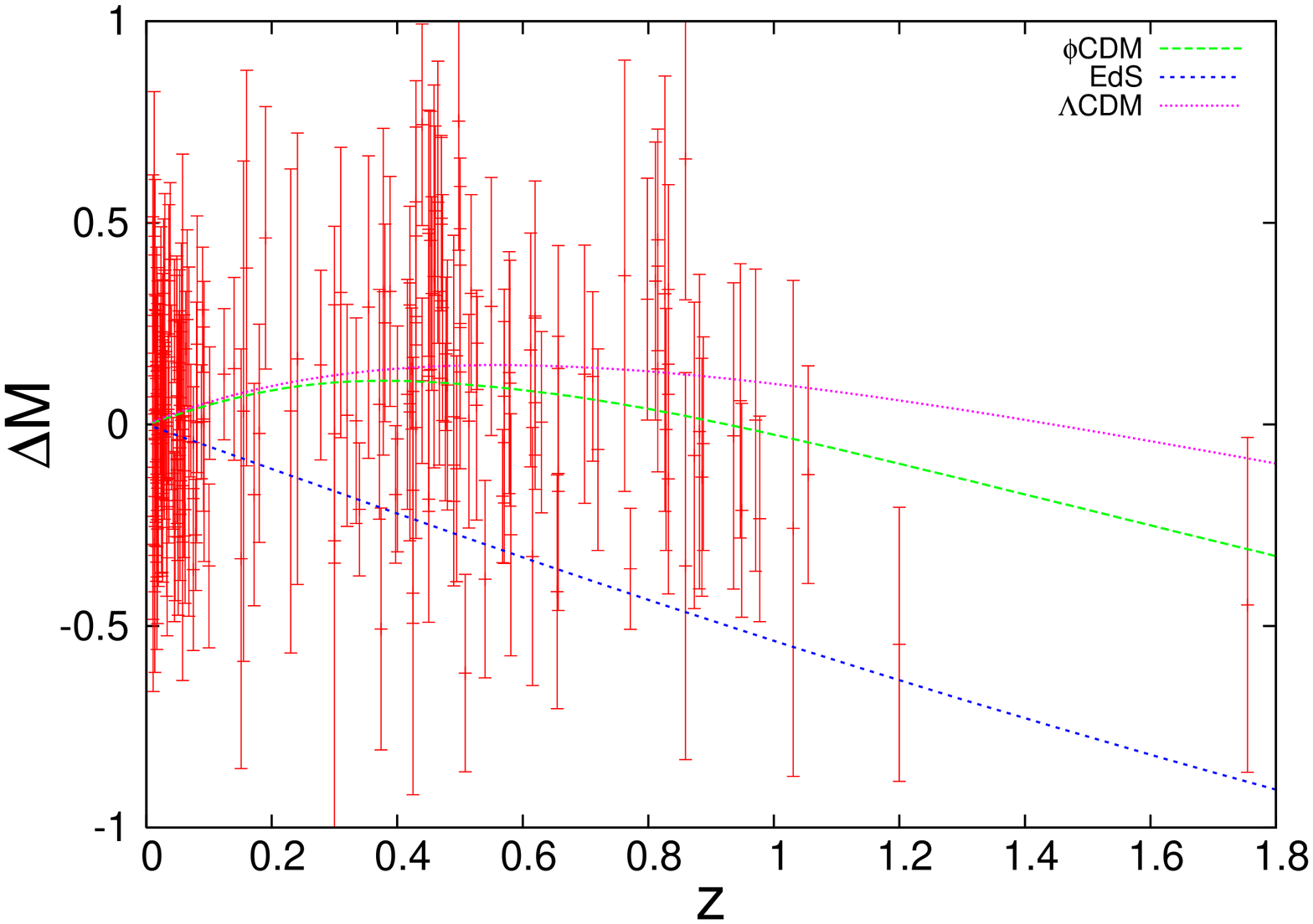}
\caption[fig4] {\label{fig4} Luminosity vs. distance with respect to
the empty universe for $(\beta=1.3,~\mu=8)$ (green, long-dashed),
$\Lambda$CDM model (purple, dot-dashed) and EdS (blue, short-dashed)
along with the SNIa data.}
\end{figure}

%%%%%%%%%%%%%%%%%%%%%%%%%%%%%%%%%%%%%%%%%%%%%%%%%%%%%%%%%%%%%%%%%%

\subsection{Observational Constraints: Supernova type Ia data}

In order to see whether the model presented here is compatible with
present observational data, we compare it with the recent expansion
history as probed by the supernova type Ia (SNIa) data.  Here we use
the sample of 194 SNIa presented in \cite{barris}.  In order to
compare our model with the \lcdm model, we compute the $\chi^2$ for
each set of parameters and compare it with the value for the
corresponding flat \lcdm value, which for the same set of supernovae
is $\chi^2_{\Lambda CDM}=201.6$
($\Omega_m=0.27,\Omega_{\Lambda}=0.73$).  The contribution from
radiation is neglected for the SNIa fit, and the universe is assumed
to be flat. The Hubble parameter is marginalized over in all of the
fits (by maximizing the likelihood).

In Fig.~\ref{fig3} we show a contour plot of coefficients
($\mu,~\beta$) corresponding to $\Delta\chi^2\equiv
\chi^2-\chi_{\Lambda CDM}^2$, where we have fixed $\Omega_{m}=0.27,\
\Omega_{sc}=0.01$.  The region above the blue (short-dashed) curve
shows where the $\chi^2$ is smaller than $\chi^2_{\Lambda CDM}$,
indicating a nominally better fit to the data.  The green and red
curves indicate contours where $\Delta\chi^2$ is $-2$ and $-4$,
respectively.  The reduced $\chi^2$ for the $\Lambda$CDM model is
$201.6/(194-1)\approx 1.045$, so an equally good reduced $\chi^2$ is
achieved with the model considered here (with two extra parameters),
when $\Delta\chi^2\approx -2.1$.

The dependence on the coupled matter parameter $\Om_{sc}$ is very
weak, whereas the uncoupled component plays a significant role. For
example, with large values, $\Om_{m}\geq 0.4$, the contours shift such
that all of the considered $(\beta,\mu)$ parameter space gives a
larger $\chi^2$ than the \lcdm value. Small values, $\Om_{m}\leq 0.2$,
lead to the same conclusion. Here our main purpose is to demonstrate
the properties of the model considered here and to show how it can fit
the current SNIa data very well.  A more complete parameter scanning
is left for future study,

In order to understand qualitatively why $\Delta\chi^2$ is negative
for a range of parameter values, we show the luminosity distance plot
as a function of redshift with respect to the empty universe, $\Delta
M$, for a set of parameters $(\beta=1.3,\ \mu=8)$, for which
$\chi^2-\chi^2_{\Lambda CDM}=-4.2$, in Fig.~\ref{fig4}.  From the
figure we see how for these parameters the luminosity distance at low
redshifts closely follows that of the $\Lambda$CDM but at later times
curves down more quickly, fitting the high redshift data points
better.

%%%%%%%%%%%%%%%%%%%%%%%%%%%%%%%%%%%%%%%%%%%%%%%%%%%%%%%%%%%%%
\section{Stringy Physics and Cosmic Acceleration:}

Although we have seen that stringy SCDM can pave the way for a late
time acceleration phase, challenges remain to connect string theory
with our coupled quintessence model in a quantitative way.  Firstly,
our numerical analysis involving supernovae data suggests that we
require $\mu/\bb\geq 2$ and $\bb\geq 1.4$ for a good fit. The
parameters are further pushed up if one also considers bounds coming
from BBN: During the radiation era $(\mu_r=0)$ we have
\be 
\ti{\rho}_\phi={5\over6(\beta^2-1)}\ti{\rho}_{r}\,.
\label{bbn}
\ee 
BBN considerations provide an upper bound on the dark energy abundance,
$\Om_\phi<0.21$ during the radiation era. From Eq.~(\ref{bbn}) we see that
for this to be satisfied one needs $\bb>2$. 

Secondly, it is clear that the success of our cosmological scenario 
relies on having a very small abundance of CDM and SCDM in the early 
universe, and it would not be much progress if we simply shift the usual
problem of quintessence models, namely the problem of having to
invoke a new and extremely small mass hierarchy (namely 
the $ev$ mass scale), to a problem of an unnaturally small
initial abundance of the SCDM particles. Thus in order to truly 
address the cosmic 
acceleration issue one needs to explain quantitatively why we expect 
such small abundances. In this section we address both of these issues, 
starting with the first. 

\subsection{Quantum Stringy Corrections}
 
One may argue that we can obtain the required values of $\mu,\bb$ if
stringy SCDM is coupled to several scalar fields
\footnote{Note that invoking several scalar fields has also been
shown to make it easier to realize a period of cosmological
inflation \cite{assisted,Ninflation,Jokinen}.}. This is because,
depending upon which linear combinations are stabilized, the effective
$\mu$ for $\phi$ (the linear combination which is rolling), can range
from $0$ to $\sqrt{\sum_I \mu_I^2}$, where $\mu_I$'s are the coupling
exponents to the different scalar fields, $\phi_I$'s. Analogous arguments 
hold for $\bb$. However, here rather than
invoking many scalar fields, we furnish an example showing how quantum
corrections involving stringy loops may be able to provide us with a
significantly large value of $\mu$.

In the string frame quantum string loop corrections will in general
modify the couplings of the dilaton to the different fields:
\be
S=\int d^{10}x~[C(\phi)R+K(\phi)(\p\phi)^2-V(\phi)]\,,
\ee 
where the functions $C(\phi), K(\phi)$
and $V(\phi)$ are expected to have a Taylor series
expansion in terms of the inverse string coupling constant
$g_s^{-1}\equiv e^{-\phi}$ \cite{veneziano}
\be
C(\phi)=\sum_{n=0}^{\infty}C_ne^{-n\phi}\,,
\ee
and similarly for $K(\phi)$ and $V(\phi)$. We recover the classical
action (\ref{typeII}) when all co-efficients, except $C_2=1$ and
$K_2=4$, vanish. Now, let us suppose that the first non-zero
coefficients of $C(\phi)$ and $K(\phi)$ occur at the $n'th$ level,
while $V_0=0$ \footnote{One could consider more general conditions, but
since this is only for illustration, we choose a simple ansatz.}. Then,
as $\phi\ra\infty$, the action is given by
\be
S\approx\int d^{10}x\ [e^{-n\phi}R+e^{-n\phi}K_n(\p\phi)^2-V_1e^{-\phi}]\,.
\ee   
To go to the Einstein frame we perform a conformal transformation
\be
\hat{g}_{\mh\nh}\ra e^{-2n\phi/(\Ds+2)}\hat{g}_{\mh\nh}\,,
\ee
followed by rescalings
\be
\phi\ra \sqrt{\frac{n^2(\Ds+3)}{\Ds+2}-K_n}\phi\equiv k\phi\,,
\label{k}
\ee
to make the kinetic terms for the dilaton canonical.
After dimensional reduction to 4 dimensions, the action reads
\be
S=\frac{1}{16\pi G}\int dx^{4}\sqrt{-g}\left[R-\p_{m}\phi\p^{m}
\phi-V_1e^{-2\bb\phi}\right]\,,
\ee
where
\be
2\bb=\frac{\Ds+2-2n}{k(\Ds+2)}\,.
\ee

Next, we look at the DBI action for $\Ds$-branes and perform the
conformal rescalings (\ref{conf-phi}) to obtain the effective 4
dimensional action for the gas of 6-branes:
\bea
S_{\mt{brane}} \, &=& \, \int d^4x \sqrt{-g}\ti{\rho} \\
&=& \int d^4x 
\sqrt{-g}\rho_0e^{-2\mu_{sc} \phi}\left(\frac{a}{a_0}\right)^{-3}\,,
\nonumber
\eea
with
\be
\mu_{sc}=\frac{\Ds(n-1)+n-2}{k(\Ds+2)}\,.
\ee
Note that we want $\mu/\bb$ to be positive.  Substituting $\Ds=6$ one
finds that this is possible only for $n=2,3,4$:
$$\mx{For }n=2,\ \frac{\mu_{sc}}{\bb}=1$$
$$\mx{For }n=3,\ \frac{\mu_{sc}}{\bb}=\frac{13}{4}\approx 3.13$$
$$\mx{For }n=4,\ \frac{\mu_{sc}}{\bb}=10$$
From Eq.~(\ref{k}) it is also clear that, depending on the value of
$K_n$, $k$ could be small, hence making $\mu_{sc},\bb$ large.

%%%%%%%%%%%%%%%%%%%%%%%%%%%%%%%%%%%%%%%%%%%%%%%%%%%%%%%%%%%%%
\subsection{Primordial Inflation and the Scale of Current Acceleration}

It is clear that a resolution of the dark energy problem should not
only involve (a) a tracking mechanism which explains why the dark
energy density has always been close to the matter density, and not
just today (and this is most certainly the case in our model), but
also (b) a parameter (since we do not want to introduce a hierarchical
$ev$ mass scale) which governs when the dark energy comes out of the
tracking phase and starts to dominate the universe. Moreover, this
parameter should be such that its expected value can naturally explain
why we are entering this phase of acceleration so late in the day.

Now, in our model, acceleration commences approximately when SCDM
becomes comparable to ordinary CDM. Since SCDM redshifts more slowly
than CDM, this means that in the early universe (say just after
inflation) the energy density in SCDM was much smaller than that of
CDM. This could happen, for example, if the SCDM is produced very
scantily during the reheating process, as argued in Section II. More
importantly, one realizes that SCDM candidates are stringy winding
states in nature.  One expects these winding states to be present in
the very early pre-inflationary universe with approximately string
scale energy densities. However, as our universe expands exponentially
during inflation, this gas of branes is going to become very
dilute. Since the slope of the potential decreases exponentially as
$\phi$ increases, we expect the dilaton $\phi$ to be effectively
frozen during the phase of inflation.  As a result, the SCDM energy
density will redshift as $a^{-3}$.

Thus, at the end of inflation, while the energy density in ordinary
matter would be given by the reheat temperature, the energy density in
the primordial gas of SCDMs would be much smaller due to inflationary
dilution. In fact, the larger the number of e-foldings ${\cal N}$, the
greater the dilution, and therefore the longer it will take for SCDM
to catch up with CDM, and the smaller would be the scale ($\equiv
M_A$) at which the second phase of acceleration commences.  If
$M_A\equiv M_p10^{-A}$, then one finds (see Appendix for details)
\be
A=\frac{R(1+4\mu/\bb-3\om_{sc})+(1.3{\cal N}-27)(1+\om_{sc})}{4(\mu/\bb-\om_{sc})}
\label{accel}
\ee
where $M_P10^{-R}$ corresponds to the reheat temperature.
 
We are now in a position to estimate the scale at which the second
acceleration phase starts. As a first example, let us choose $\om_{sc}=0$,
and $\mu/\bb=1/2$, the minimal value required for acceleration and
also approximately what one obtains for the 6-branes. For the minimal
number of e-foldings ${\cal N}=60$ and assuming reheating to GUT
scale temperatures $M_{reh}$, i.e. $M_{reh}\sim 10^{-3}M_p$ or $R=3$, one
finds from (\ref{accel}) that $M_A\sim 10^{-30}M_p=10^{-2}\ ev$. We remind
the reader that the current energy density corresponds to the energy scale
$10^{-3}\ ev$. For $\om_{sc}=1/3$, i.e. radiative SCDM, one needs
$\mu/\bb=1$ for acceleration. Then using previous values for ${\cal
N}$ and $R$ we again find $M_A\sim 10^{-30}M_p$. 

The agreement, however, is not as dramatic as it seems. The more
detailed numerical analysis performed with $\om_{sc}=0$ revealed that
we get good agreement with the supernova data for values of
$\mu/\bb\geq 2$. Demanding $A\sim 31$, and $\mu/\bb\sim 2$ one
``predicts'' the number of primordial inflation to be ${\cal N}\sim
200$ if one demands that the second stage of acceleration is beginning
at the present time.  Estimates with $\om_{sc}=1/3$, and $\mu/\bb\sim
5$ (\ref{1-ratio}) yield a similar estimate for ${\cal N}$, namely
${\cal N}\sim 250$. Of course, this number depends on the reheat
temperature, and on the precise pre-inflationary initial energy
density of SCDM, but these dependences are relatively weak as can be
seen from (\ref{accel}). Thus, (\ref{accel}) should be treated as a
relation between $A$ and ${\cal N}$. Moreover, since $A=A({\cal N})$
is a linear relationship, there is no fine tuning involved; a small
variation in ${\cal N}$ results in only a small variation in $A$.
This is of course to be contrasted with the fine tuning problem in
$\La$CDM models where the difference of two large numbers is supposed
to yield a very small cosmological constant. Hence, a small variation
in one of the large numbers causes a huge variation in $\La$.

%%%%%%%%%%%%%

\section{Conclusions}

In this paper we have introduced the concept of ``stringy cold dark
matter'' (SCDM). From the point of view of our four-dimensional
space-time, SCDM are particles which couple non-trivially (with a
coupling $exp(2 \mu \phi)$) to the dilaton. Candidates for SCDM from
string theory are branes wrapping some or all of the compact spatial
dimensions. We have studied the cosmology of these modes in the
context of a Type IIA supergravity action to which we have added a
run-away potential $V_0 exp(- 2 \beta \phi)$ for the dilaton $\phi$
(as explained in the text, such potentials are generally believed to
be generated when going beyond the classical description). We have
shown the existence of cosmological tracking solutions in the context
of which the dilaton becomes a candidate for dark
energy~\footnote{Cosmological tracking solutions from exponential
potentials have been also considered in \cite{Copeland} and references
therein.}.

Our model contains radiation, regular dark matter (which couples very
weakly if at all to the dilaton), SCDM and the dilaton. Assuming that
our four-dimensional space-time underwent a period of primordial
inflation, the initial energy density in SCDM is expected to be
exponentially suppressed compared to the density of regular CDM and
radiation (which can be produced during reheating after inflation). In
this context, the universe is initially dominated by radiation, and we
show that the energy density in the dilaton tracks that of radiation
until the dilaton gets stuck at the minimum of its effective
potential, a potential formed from the dilaton potential (potential
with a negative slope) and the terms with positive slope coming from
the interactions of the dilaton with the SCDM. From then on, the
density in SCDM and in the dilaton redshift less fast than that of
ordinary matter. During this phase, the density in the dilaton is
greater than that in SCDM. Once the dilaton energy begins to dominate,
a period of acceleration begins, provided $\mu / \beta$ is
sufficiently large.

Compared to models of quintessence, an advantage of our approach is
that it does not require the introduction of a new mass hierarchy (a
new mass scale of about $1$eV). In order to explain why the late-time
acceleration begins at the present time, a sufficient suppression of
the number density of SCDM states is required. Such a suppression is
naturally generated by a period of cosmological inflation. In fact,
the number ${\cal N}$ of e-foldings of inflation required to suppress
the SCDM density is (for GUT-scale reheating) larger but of the same
order of magnitude as the minimal value of ${\cal N}$ required to
successful inflation.  Thus, in our framework the ``coincidence
mystery'' of why dark energy is beginning to dominate today is tied to
the duration of the period of inflation.

For our scenario to work, a sufficiently large value of $\mu / \beta$
is required. Even larger values of this ratio are required to be
consistent with big bang nucleosynthesis.  At the level of the
classical action, it is not possible to obtain such large values with SCDM particle coupling to a single scalar moduli.  The cumulative effect of SCDM coupling to several different scalar fields may help resolve this difficulty, as may non-perturbative effects.  Further research on this issue is required.

It would be interesting to study further consequences of the existence
of SCDM in the present universe, following the approaches of
\cite{Gubser2} to study cosmological consequences and of \cite{shiu}
to study astrophysical and particle physics aspects.

%%%%%%%%%%%%%%%%%%%%%%%%%%%%%%%%%%%%%%%%%%
\section{Acknowledgments}
T.~B. is supported by the NSERC Grant No. 204540. The work of R.~B. is
supported by an NSERC Discovery grant and by the Canada Research Chair
Program.

%%%%%%%%%%%%%%%%%%%%%%%%%%%%%%%%%%%%%%%%%%
\section{Appendix}

We start by assuming that in the very early universe the energy
density of the stringy SCDM and of $\phi$ is given by the string
scale, $\ti{\rho}_{sc}\sim V(Q)\sim M_{str}^4$. Now, a small hierarchy
between the string scale and the scale of inflation ensures that $Q$
or SCDM play no role in the inflationary mechanism and that $\phi$ is
effectively frozen during this phase. As a result, the SCDM energy
density redshifts as $a^{-3(1+\om_{sc})}$ with the rapid expansion of
our universe. Accordingly, after the end of inflation the SCDM energy
density will be given by
\be
\ti{\rho}_{sc}(t_R) \, = \, \rho_I e^{-3(1+\om_{sc}){\cal N}}
\label{scmE}
\ee
where ${\cal N}$ is the number of e-foldings of inflation, and
$t_R$ stands for the time of the end of inflation (the time of
reheating), and $\rho_I$ is the energy density at the beginning
of inflation (when the inflaton potential becomes dominant). It is
a good assumption that the SCDM density will not be suppressed
relative to that of other matter between the initial time and the
onset of inflation.  

We choose, by convention, $\phi(t_R) = 0$ and thus 
$V(\phi)= M_{str}^4e^{-2\bb \phi}$, or $V_0=M_{str}^4$. Let us also assume, 
just for the purpose of illustration, that reheating is very fast, i.e. 
that the period of inflation is immediately followed by a radiation
period (this is likely to be the case of reheating is driven by
parametric resonance). In this case, after reheating, the radiation energy 
density is given by
\be
\ti{\rho}_{r}(t_R) \, = \, \rho_I \equiv M_p^410^{-4R} \, ,
\label{radE}
\ee
where $R$ sets the scale of inflation (which is $10^{-R} M_p$).
It is now clear from (\ref{scmE}) and (\ref{radE}) that after the 
end of inflation for ${\cal N}\geq 60$, and $\om_{sc}\geq 0$, the 
radiation energy density is a lot larger than the SCDM density.
This is essentially why it takes a long time for $\ti{\rho}_{sc}$ to catch up 
with radiation density. As we have seen, once it does, a second  phase of 
accelerated expansion begins. Let us therefore try to estimate when 
SCDM becomes comparable to ordinary matter/radiation. 

After inflation, all the scalar fields are free to roll. The energy 
density of the $\phi$ field will first track the radiation and then 
matter energy density, maintaining a constant ratio with them. Thus, in 
the radiation era
\be
e^{-2\bb \phi} \, \sim \, a^{-4} \, 
\Ra \, e^{2\mu \phi} \, \sim \, a^{4\mu/\bb}
\ee
Hence the radiation and SCDM densities redshift differently
\bea
\ti{\rho}_{sc} \, &=& \, \rho_I e^{-3(1+\om_{sc}){\cal N}}
\left(\frac{a}{a_E}\right)^{4\mu/\bb-3(1+\om_{sc})} \nonumber \\
&=& \, M_P^4 10^{-1.3(1+\om_{sc}){\cal N}-4R}
\left(\frac{a}{a_E}\right)^{4\mu/\bb-3(1+\om_{sc})}
\label{scm}
\eea
and
\be
\ti{\rho}_{r} \, = \, M_p^410^{-4R}\left(\frac{a}{a_E}\right)^{-4} \, .
\label{radiation}
\ee

After the radiation-matter equality, when the energy density of the 
universe has fallen to about $10^{-108}M_p^4$, $\phi$ starts to track 
the matter density instead
\be
e^{-2\bb \phi} \, \sim \, a^{-3}\, \Ra \, e^{2\mu \phi}
\, \sim \, a^{3\mu/\bb}
\ee
Now, one can obtain the ratio $a_{eq}/a_R$, where ``$t_{eq}$'' is the
time of radiation-matter equality and $a_{eq}$ is the value of the
scale factor at that time, and then substitute it in (\ref{scm}) to
obtain the energy density of SCDM at the equality epoch:
\bea
\ti{\rho}_{sc}((t_{eq}) \, &=& \, M_p^410^{-1.3(1+\om_{sc}){\cal N}-4R}
\nonumber \\
&& 10^{(27-R)(4\mu/\bb-3(1+\om_{sc}))} \, .
\eea
The evolution of strongly coupled and ordinary matter is then given by
\be
\ti{\rho}_{sc}=\ti{\rho}_{sc}(t_{eq})\left(\frac{a}{a_{eq}}\right)^{3\mu/\bb-3(1+\om_{sc})}
\label{scm2}
\ee
and
\be
\ti{\rho}_{m}=M_p^410^{-108}\left(\frac{a}{a_{eq}}\right)^{-3} \, ,
\label{matter}
\ee
respectively.

Now, as explained before, $\ti{\rho}_{sc}\sim\ti{\rho}_{m}\equiv M^4_A$ 
corresponds to the acceleration epoch. If $M_A\equiv M_p10^{-A}$, 
then from (\ref{matter}) and (\ref{scm2}) we have
\be
A=\frac{R(1+4\mu/\bb-3\om_{sc})+(1.3{\cal N}-27)(1+\om_{sc})}{4(\mu/\bb-\om_{sc})}
\label{accel}
\ee

%%%%%%%%%%%%%%%%%%%%%%%%%%%%%%%%%

%%%%%%%%%%%%%%%%%%%%%%%%%%%%%%%%%

\end{document}